\documentclass{ws-procs9x6}

\begin{document}

\title{Theoretical Aspects of High Energy Neutrinos and 
GRB\footnote{\uppercase{B}ased on the talk given by \uppercase{P.M.}
in the \uppercase{I}nternational \uppercase{W}orkshop on
\uppercase{E}nergy \uppercase{B}udget in the \uppercase{H}igh 
\uppercase{E}nergy \uppercase{U}niverse, \uppercase{K}ashiwa, 
\uppercase{J}apan, \uppercase{F}ebruary 2006.}}

\author{P. M\'esz\'aros and S. Razzaque}

\address{Department of Astronomy and Astrophysics,
Department of Physics, \\ Pennsylvania State University, University
Park, PA 16802, USA \\ E-mail: nnp@astro.psu.edu, soeb@astro.psu.edu}

\maketitle

\abstracts{Abstract: Neutrinos at energies ranging from sub-TeV to EeV
from astrophysical sources can yield interesting physical information
about fundamental interactions, about cosmic rays and about the nature 
of the sources and their environment. Gamma-ray bursts are a leading 
candidate source, and their expected neutrino emission can address a 
number of current questions, which may be answered with forthcoming
experiments such as IceCube, Auger, ANITA and KM3NeT.}

\section{Introduction}

The origin of the observed ultrahigh-energy (UHE) cosmic-rays (CRs) above
the ``ankle'', roughly at EeV ($= 10^{18}$ eV) energy, of the CR
energy spectrum is most probably extra-galactic. Any galactic origin
at this energy, due to small magnetic deflections, would result in
an anisotropic distribution of their arrival direction contrary to
the observed data. The requirement that they are not attenuated by the
cosmic microwave background through photo-meson ($p\gamma$)
interactions constrains them to have originated within a radius of 
50-100 Mpc, the so-called ``GZK'' volume.\cite{g66,zk66} 
Two broad classes of models suggested are the ``top-down'' scenarios, 
which attribute UHECRs to the decay of fossil Grand Unification defects, 
and the ``bottom-up'' scenarios, which assume UHECRs are accelerated in 
astrophysical sources.

The observed UHECR energy injection rate into the universe is $\sim
3\times 10^{44} ~{\rm erg~Mpc^{-3}~yr^{-1}}$ above the ankle. This is
similar to the $\sim 0.1$-$1$ MeV $\gamma$-ray energy injection rate
by the local gamma-ray bursts (GRBs) which led to postulating that
GRBs are the sources of UHECRs.\cite{v95,w95a} This coincidence has
been corroborated using new data and further
considerations,\cite{wax04,vdg03,wick04} making GRBs promising
candidates for UHECRs.  Other candidates, in the bottom-up scenario,
are active galactic nuclei (AGNs), and cluster accretion shocks. An
unavoidable by-product of UHECR acceleration is the production of UHE
neutrinos, via $p\gamma$ and $pp,~pn$ interactions.  We limit our
discussion to UHE neutrinos from GRBs here.

\section{Nature of the High Energy Emission from GRBs}

In the most widely accepted GRB model, the {\em fireball shock model},
the prompt $\gamma$-rays are produced by shocks in the plasma material
ejected in a jet moving relativistically (with a bulk Lorentz factor
$\Gamma \gtrsim 100$), usually taken to be ({\em internal shocks}), or
in other versions an {\it external shock} (see, e.g.,
Ref. [\refcite{mesz02}]).  Such jets can arise from the core collapse
of massive stars, convincingly shown to be the progenitor of long
GRBs, or from mergers of compact binary systems (neutron star-neutron
star, black hole-neutron star), which may be implicated in producing
short GRBs. Late time collision of the jet material with an external
medium (external shocks) produce a long lasting x-ray, UV and optical
radiation, collectively known as the GRB afterglow.

The highly relativistic nature of the outflows is inferred from and
constrained by the observations of GeV photons which avoid attenuation
by $\gamma\gamma \to e^\pm$ production {\em in situ}.  The probable
mechanism(s) responsible for the observed photons is/are synchrotron
radiation or/and inverse Compton (IC) scattering by high energy
electrons. These electrons are accelerated by the relativistic shocks via
the Fermi mechanism in the tangled magnetic field, resulting in a power-law
energy distribution. The high bulk Lorentz factors result in synchrotron 
spectra which in the observer frame extend beyond 100 MeV, and IC scattering
of such synchrotron photons leads to the expectation of GeV and TeV
spectral components.\cite{mrp94} While $\lesssim 18$ GeV photons have
been observed,\cite{hur94} TeV photons are likely to be degraded to
lower energies by $\gamma\gamma$ pair production, either in the source
itself,\cite{razmz04_gev} or (unless the GRB is at very low redshifts)
in the intervening intergalactic medium.\cite{coppi97,dejste02}

GRBs are likely to be more luminous in neutrinos, gravitational waves
and cosmic rays compared to sub-GeV electromagnetic channels which
comprise a small fraction of the burst kinetic energy. A significant
amount of baryons (neutrons and protons) are expected to be present in
the GRB jet along with leptons, each with $\Gamma m_pc^2 \gtrsim 100$
GeV bulk kinetic energy in the observer frame.  Protons are also
expected to co-accelerate with electrons in the internal and external
shocks by the same Fermi mechanism. Using the shock parameters
inferred from broad-band photon spectral fits, one infers that protons
can be accelerated to Lorentz factors up to $\lesssim 10^{11}$ in the
observer frame, i.e. to the GZK energy of $E_p\sim 10^{20}$
eV.

\section{High Energy Neutrinos}

High energy neutrinos, detectable by the neutrino telescopes such as
IceCube in the $\sim 100$ GeV-EeV range, are produced in the GRBs in a
way similar to the beam-dump experiments in particle accelerators.
Shock-accelerated protons interacting with ambient radiation and/or
plasma material by photonuclear ($p\gamma$) and/or inelastic nuclear
($pp/pn$) collisions produce charged pions ($\pi^\pm$) and neutral
pions ($\pi^0$). Neutrinos are produced from $\pi^\pm$ decays along
with muons and electrons.

Such neutrinos may serve as diagnostics of the presence of
relativistic shocks, and as probes of the acceleration mechanism and
the magnetic field strength. The flux and spectrum of EeV neutrinos
depends on the density of the surrounding gas, while the TeV-PeV
neutrinos depend on the fireball Lorentz factor. Hence, the detection
of very high energy neutrinos would provide crucial constraints on the
fireball parameters and GRB environment. Lower energy ($\lesssim$TeV)
neutrinos originating from sub-stellar shocks, on the other hand, may
provide useful information on the GRB progenitor.

\subsection{Neutrinos contemporaneous with the gamma-rays}

With an initial $\Gamma_i = 300 ~\Gamma_{300}$ and a variability time
scale $\delta t = 10^{-3} \delta t_{-3}$ s, internal shocks in the GRB
jet take place at a radius $r_i \sim 2 \Gamma_i^2 c \delta t
\sim 5 \times 10^{12} \delta t_{-3} \Gamma_{300}^2$ cm. The fireball
becomes optically thin at a radius $\lesssim r_i$ allowing observed
$\gamma$-ray emission. Shock accelerated protons interact dominantly
with observed synchrotron photons with $\sim$MeV peak energy in the
fireball to produce a $\Delta^+$ resonance as $p\gamma \rightarrow
\Delta^+$. The threshold condition to produce a $\Delta^+$ is $E_p
E_{\gamma} = 0.2 \Gamma_i^2$ GeV$^2$ in the observer frame, which
corresponds to a proton energy of $E_p = 1.8 \times 10^{7}
E_{\gamma, {\rm MeV}}^{-1} \Gamma_{300}^{2}$ GeV. The short-lived
$\Delta^+$ decays either to $p\pi^0$ or to $n \pi^+ \rightarrow n
\mu^+ \nu_{\mu} \rightarrow n e^+ \nu_e {\bar \nu}_{\mu} \nu_{\mu}$
with roughly equal probability. It is the latter process that
produces high energy neutrinos in the GRB fireball, contemporaneous
with the $\gamma$-rays.\cite{wb97a} The secondary $\pi^+$ receive
$\sim 20\%$ of the proton energy in such an $p\gamma$ interaction
and each secondary lepton roughly shares 1/4 of the pion energy.
Thus each flavor ($\nu_e$, ${\bar \nu}_{\mu}$ and $\nu_{\mu}$) of
neutrino is emitted with $\sim 5\%$ of the proton energy, dominantly
in the PeV ($=10^{15}$ eV) range, with equal ratios.

\begin{figure}[ht]
\centerline{\epsfxsize=4.1in \epsfbox{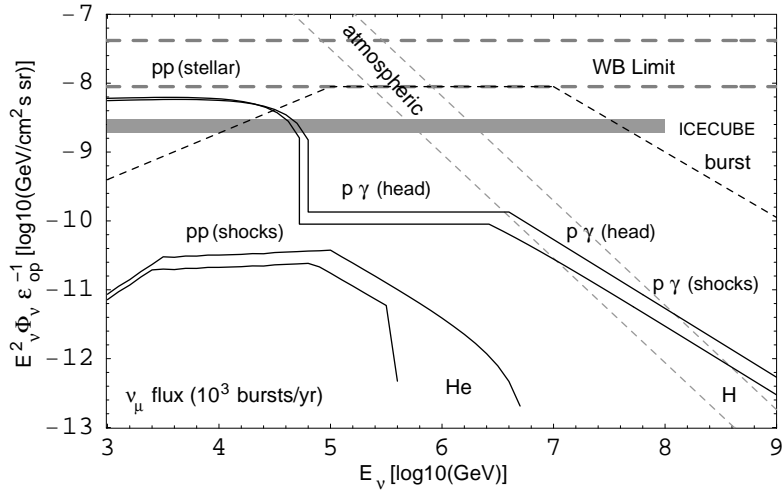}}
\caption{ Diffuse $\nu_{\mu}$ flux arriving simultaneously with the
$\gamma$-rays from shocks outside the stellar surface in observed GRB
(dark short-dashed curve), compared to the Waxman-Bahcall (WB) diffuse
cosmic ray bound (light long-dashed curves) and the atmospheric
neutrino flux (light short-dashed curves). Also shown is the diffuse
muon neutrino precursor flux (solid lines) from sub-stellar jet shocks
in two GRB progenitor models, with stellar radii $r_{12.5}$ (H) and
$r_{11}$ (He). These neutrinos arrive 10-100 s before the
$\gamma$-rays from electromagnetically detected bursts (with similar
curves for $\nu_{\mu}$, $\nu_e$ and $\nu_{\tau}$). }
\label{fig:nujet}
\end{figure}

The diffuse muon neutrino flux from GRB internal shocks due to proton
acceleration and subsequent $p\gamma$ interactions is shown as the
short dashed line in Fig. \ref{fig:nujet}. The flux is compared to the
Waxman-Bahcall limit of cosmic neutrinos from optically thin sources,
which is derived from the observed cosmic ray flux.\cite{wb00} The
fluxes of all three neutrino flavors ($\nu_e$, $\nu_{\mu}$ and
$\nu_{\tau}$) are expected to be equal after oscillation in vacuum
over astrophysical distances.

\subsection{Neutrinos from the GRB afterglow}

The GRB afterglow arises when relativistic jetted plasma material
starts being slowed down by the external medium (e.g. the interstellar 
medium, or ISM), driving a blast wave ahead of the jet. This produces 
an external forward shock or blast wave, and a reverse shock in the jet.
The external shock takes place at a radius $r_e \sim 4\Gamma_e^2 c
\Delta t \sim 2 \times 10^{17} \Gamma_{250}^2 \Delta t_{30}$ cm which
is well beyond the internal shock radius.\cite{wb00} Here $\Gamma_{e}
\approx 250 \Gamma_{250}$ is the bulk Lorentz factor of the ejecta
after the partial energy loss from emitting $\gamma$-rays in the internal
shocks, and $\Delta t = 30 \Delta t_{30}$ s is the duration of the GRB
jet. Neutrinos are produced in the external reverse shock due to
$p\gamma$ interactions of shock accelerated protons predominantly with
synchrotron soft x-ray photons produced by electrons. The energy of
the neutrinos from the afterglow would be in the EeV range as more
energetic protons interact with these soft photons to produce
$\Delta^+$. The efficiency of proton to pion conversion by $p\gamma$
interactions in the external shocks (afterglow) is typically smaller
than in the internal shocks because $r_e \gg r_i$, implying lower
photon density.

In the case of a massive star progenitor the GRB jet may be expanding
into a stellar wind much denser than the typical ISM density of $n\simeq 
1~{\rm cm}^{-3}$, which is emitted by the progenitor prior to its collapse. 
For a wind with mass loss rate of $\sim 10^{-5} M_\odot~{\rm yr}^{-1}$ and
velocity of $v_w\sim 10^3~{\rm km/s}$, the wind density at the typical
external shock radius would be $\simeq 10^4~{\rm cm}^{-3}$. The higher
density implies a lower $\Gamma_e$, and hence a larger fraction of
proton energy lost to pion production. Protons of energy $E_p\gtrsim
10^{18}$~eV lose all their energy to pion production in this scenario
producing EeV neutrinos.\cite{dl01}

\subsection{Precursor neutrinos}

In the long duration GRBs, the relativistic jet is expected to be launched 
near the central black hole  resulting from the collapse of the stellar core, 
hence the jet is initially buried deep inside the star. As the jet burrows 
through the stellar material, it may or may not break through the stellar
envelope.\cite{mw01} Internal shocks in the jet, while it is burrowing
through the stellar interior, can produce high energy neutrinos due to
accelerated protons, dominantly below $\sim 10$ TeV, through $pp$ and 
$p\gamma$ interactions.\cite{rmw03b} The jets which successfully penetrate
through the stellar envelope result in GRBs ($\gamma$-ray bright bursts), 
while the jets which choke inside the stars do not produce GRBs
($\gamma$-ray dark bursts). However, in both cases high energy
neutrinos can be produced in the internal shocks, which slice through 
the stellar envelope since they interact very weakly with matter.

These neutrinos from the relativistic buried jets are emitted as
precursors ($\sim$ 10-100 s prior) to the neutrinos emitted from the
GRB fireball in case of an electromagnetically observed burst. In the
the case of a choked burst (electromagnetically undetectable) no
direct detection of neutrinos from individual sources is possible.
However the diffuse neutrino signal is boosted up in both scenarios.
The diffuse neutrino flux from two progenitor star models are shown in
Fig. \ref{fig:nujet}, one for a blue super-giant (labeled H) of radius
$R_\ast=3 \times 10^{12}$ cm and the other a Wolf-Rayet type (labeled
He) of radius $R_\ast =10^{11}$ cm. The Waxman-Bahcall diffuse cosmic
ray bound,\cite{wb99} the atmospheric flux and the IceCube sensitivity
to diffuse flux are also plotted for comparison.  The neutrino
component which is contemporaneous with the gamma-ray emission
(i.e. which arrives after the precursor) is shown as the dark dashed
curve, and is plotted assuming that protons lose all their energy to
pions in $p\gamma$ interactions in internal shocks.

\subsection{Early $np$ decoupling non-thermal neutrinos}

Neutrons are expected to be present in considerable numbers in the GRB
jet ($n_n \simeq n_p$) because of a {\em neutronized} core similar to
that in supernovae in the case of long GRB, and from neutron star 
material in the case of a short GRB. In the long GRB, the core collapse 
neutronization leads to copious thermal ($\sim 10$ MeV) neutrinos, but
due to their low energy, their cross section is too small for detection 
at cosmological distances. However, in both long and short GRB outflows,
neutrons are present, and are initially coupled to the protons by elastic 
nuclear scattering. If the initial acceleration of the fireball is very 
high, the neutrons can eventually decouple from the fireball, when
the comoving expansion time falls below the nuclear scattering time. 
Protons, on the other hand, continue accelerating and expanding with 
the fireball as they are coupled to the electrons by Coulomb scattering. 
The relative velocity between the protons and neutrons, in such a case, 
can get high enough for inelastic interactions ($np$) above the pion 
production threshold of $\sim 140$ MeV, leading to $\sim 10$ GeV neutrinos 
in the observer's frame.\cite{dkk99,bm00,mr00b} Highly neutron-enriched
($n_n \sim 10 n_p$) jet in case of short GRBs may lead to $\sim 50$
GeV neutrinos, as the relative velocity between the protons and
neutrons increases substantially, which are detectable from a nearby
burst.\cite{rm06a}

\section{GRB-Supernova Connection}

A fraction of long GRBs have recently been shown to be associated with
supernovae of type Ib/c.\cite{DellaValle:2005cr} A GRB jet loaded with
baryons would then leave long-lasting UHE CR, neutrino and photon
signatures in those supernova remnants which were associated with a GRB
at the time of their explosion. One example may be the SN remnant W49B
which is probably a GRB remnant. A signature of a neutron component in
the relativistic jet outflow would be a TeV $\gamma$-ray signature due
to inverse Compton interactions following neutron decay.\cite{iokkm04}
Another example may be some of the HESS unidentified sources.\cite{atoy05} 
Neutron decay would also give rise to TeV neutrinos. The imaging of the 
surrounding emission could provide new constraints on the jet structure 
of the GRB.

Cosmic-rays accelerated in the GRB remnant, similar to SN remnants
which are observed as TeV $\gamma$-ray sources such as RX
J1713.7-3946, would also be expected to produce UHE
neutrinos.\cite{ah02} Expected neutrino and $\gamma$-ray energy,
commonly originating from $p\gamma$ and/or $pp/pn$ interactions, would
be higher in case of GRB remnants because of the higher expansion
velocity.

\section{Neutrino Flavor Astrophysics}

High energy neutrinos from astrophysical optically thin sources are 
expected to be produced dominantly via $p\gamma$ interactions. Subsequent 
decay of $\pi^+$ and neutrino flavor oscillations in vacuum lead to an 
observed anti-electron to total neutrino flux ratio of $\Phi_{\bar\nu_e}:
\Phi_{\nu} \simeq 1:15$.\cite{lp95} At high energy 
this ratio may be lower even,\cite{kw05} since the muons suffer
significant electromagnetic energy loss prior to decay.\cite{rachen98}
In the case of $pp/pn$ interactions, typically attributed to optically
thick sources, $\pi^\pm$ are produced in pairs and the corresponding
expected flux ratio on Earth is $\Phi_{\bar\nu_e}: \Phi_{\nu} \simeq
1:6$. However even in the optically thin sources the nominal
$\Phi_{\bar\nu_e}: \Phi_{\nu}$ ratio may be enhanced above $1:15$ by
$\gamma\gamma\to \mu^\pm$ interactions and subsequent $\mu^\pm$
decays.\cite{rmw06a} The targets are usual synchrotron photons and UHE
incident photons are provided by the $p\gamma\to p\pi^0 \to
p\gamma\gamma$ channel itself. This mechanism yields an enhancement
ratio $\Phi_{\bar\nu_e}: \Phi_{\nu} \simeq 1:5$ solely from $\mu^\pm$
decays.

\begin{figure} [ht]
\centerline{\epsfxsize=4.1in \epsfbox{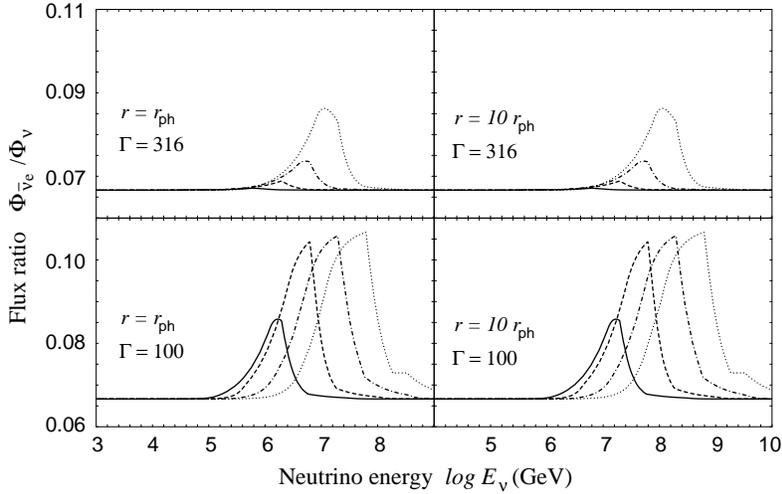}} 
\caption{ Expected anti-electron to total neutrino flux ratio: 
$\Phi_{\nu_e}/ \Phi_{\nu}$ on Earth from a GRB after vacuum
oscillations as function of neutrino energy. The fluxes are both from
the $p\gamma \to n\pi^+$ and $\gamma\gamma \to
\mu^+\mu^-$ interactions. Depending on the GRB model parameters such
as the internal shock radius (at the photosphere $r_{\rm ph}$ and at
$10r_{\rm ph}$), bulk Lorentz factor ($\Gamma$) and magnetization
($\varepsilon_B = 10^{-1},~ 10^{-2},~ 10^{-3}$ and $10^{-4}$ denoted
by solid, dashed, dotted-dash and dotted lines respectively), the flux
ratio may be enhanced from the nominal $1/15$ value in certain energy
ranges. }
\label{fig:fluxratio}
\end{figure}

Measurement of the ${\bar\nu_e}$ to $\nu$ flux ratios may be possible
by IceCube at the Glashow resonant interaction ${\bar \nu}_e e \to W^-
\to {\rm anything}$ at $E_{\nu} \simeq 6.4$
PeV.\cite{aghw05} Any enhancement over the $1:15$ ratio, e.g., from a
single nearby GRB would then suggest a $\gamma\gamma$ origin. However,
the flux of $\gamma\gamma$ neutrinos depend on the source model such
as magnetization, radius etc. We have plotted the ${\bar\nu_e}$ to
$\nu$ flux ratio in Fig. \ref{fig:fluxratio}, which includes the
contribution from $p\gamma$ and $\gamma\gamma$ channels, from a GRB
internal shocks with different model parameters. The solid, dashed,
dotted-dash and dotted lines correspond to the magnetization parameter
$\varepsilon_B= 10^{-1}$, $10^{-2}$, $10^{-3}$ and $10^{-4}$
respectively. The shocks take place at the photosphere ($r_{\rm ph}$)
and at a radius $10r_{\rm ph}$.\cite{rmw06a} Note that the ratio is
enhanced from the $p\gamma$ value of 1/15 in the small energy range
where $\gamma\gamma$ interactions contribute significantly. This
result then may be used to learn about the GRB model parameters.

\section{Conclusions}

Although fireball shock model is the leading GRB scenario, there is no
strong direct proof so far for the internal shock or the reverse shock 
origin of the observed radiation. High energy neutrino emission from GRBs 
would serve as a direct test for this, as well as for ``baryonic" jet models, 
where the bulk of the energy is carried by baryons. On the other hand, an 
alternative Poynting flux dominated GRB jet model would have to rely on 
magnetic dissipation and reconnection, accelerating electrons and hence 
also accelerating protons-- but there would be much fewer protons to 
accelerate and probably to much lower energy.

The {\em Pierre Auger Observatory}, a CR detector currently under
construction, will have very large ($\sim 3000$ km$^2$ each for its
two location in the Southern and Northern hemisphere)
area.\cite{auger_site} It will help to disentangle the two scenarios
(top-down or bottom-up) and will reveal whether a GZK feature indeed
exists by greatly improving the UHECR count statistics. Within the
bottom-up scenario, the directional information may either prove or
significantly constrain the alternative AGN scenario, and may
eventually shed light on whether GRBs are indeed the sources of
UHECRs.

Upcoming experiments such as IceCube,\cite{icecube_site}
ANITA,\cite{anita_site} KM3NeT,\cite{km3net_site} and
Auger\cite{auger_site} are currently being built to detect high energy
astrophysical neutrinos. They can provide very useful information on the 
particle acceleration, radiation mechanism and magnetic fields, as well as
about the sources and their progenitors. Direct confirmation of a GRB 
origin of UHECRs is difficult but the highest energy neutrinos may 
indirectly serve that purpose pointing directly back to their sources. 
Most GRBs are located at cosmological distances (with redshift $z\sim 1$) and
individual detection of them by km scale neutrino telescopes may not
be possible. The diffuse neutrino flux is then dominated by a few
nearby bursts. The likeliest prospect for UHE $\nu$ detection is from
these nearby GRBs in correlation with electromagnetic detection.

The prospect for high energy neutrino astrophysics is very exciting,
with AMANDA already providing useful limits on the diffuse flux from
GRBs\cite{sta04_amanda,becker06_amanda} and with
IceCube\cite{ahrens04,hulth06_icecube} on its way. The detection of
TeV and higher energy neutrinos from GRBs would be of great importance
for understanding the astrophysics of these sources such as the
hadronic vs. the magnetohydrodynamic composition of the jets, as well
as the CR acceleration mechanisms involved. High energy neutrinos from
GRBs may also serve as probes of the highest redshift generation of
star formation in the Universe, since they can travel un-attenuated,
compared to the conventional electromagnetic astronomical probes.

\section*{Acknowldgements}

Work supported by NSF grant AST0307376 and NASA grant NAG5-13286.

\end{document}